\documentclass[12pt]{article}
\usepackage{epsfig}
\usepackage{amsfonts}
\usepackage{amsopn}
\usepackage{amssymb}
\usepackage{amsmath}
\usepackage{graphicx}
\usepackage{amsthm}
\usepackage{authblk}

\title{Diffeomorphism Algebra Structure and Membrane Theory}

\author[*,$\#$,$\sim$]{Jens Hoppe}
\author[$\#$,$\S$]{ Mariusz Hynek }

\affil[*]{Korea Institute for Advanced Study}
\affil[$\#$]{Royal Institute of Technology}
\affil[$\sim$]{Sogang University}
\begin{document}
\date{}
\maketitle
\thispagestyle{empty}
\let\oldthefootnote\thefootnote
\renewcommand{\thefootnote}{\fnsymbol{footnote}}
\footnotetext[4]{mkhynek@kth.se}
\let\thefootnote\oldthefootnote
\abstract{Explicit structure constants are calculated for certain Lie algebras of vectorfields on 2-dimensional compact manifolds.}

\section*{ }
A central feature of M-brane theories 
is their diffeomorphism invariance, which in a partially gauge-fixed light-cone description (leading to a polynomial Hamiltonian and a Yang-Mills type matrix model for membranes \cite{phd,gold1})\footnote{see e.g. \cite{wt,matrixmodel} for reviews} is reduced to volume preserving diffeomorphisms (VPD), i.e. diffeomorphisms (of the parameter space of the M-dimensional extended object) that have unit Jacobian, i.e. generated by divergence-free vector fields. Interestingly, volume non-preserving diffeomorphisms (VNPD) (re)appear in this partially gauge-fixed description \cite{phd,gold1,matrixmodel} in the reconstruction of the longitudinal field $\zeta$ that (apart from the  zero mode $\zeta_{0}$ and its conjugate degree of freedom $\eta$) disappears in the Hamiltonian formulation. The relativistically invariant internal mass-squared involves only the (internal) transversal coordinates $\vec{x}_{\alpha \neq 0}$ and their conjugate momenta $\vec{p}_{\alpha}:=\int{\vec{p}Y_{\alpha}(\varphi)d^{M}\varphi}$ (the $Y_{\alpha}$, together with $Y_0=1$, being a complete, orthonormal set of functions on the parameter manifold $\Sigma_{0}^{M}$, conveniently chosen as Laplace-eigenfunctions, $\Delta Y_{\alpha}=-\mu_{\alpha} Y_{\alpha}$):
\begin{equation}
\mathbb{M}^2=\sum_{\alpha=1}^{\infty}{\vec{p}_{\alpha}\cdot \vec{p}_{\alpha}+\frac{1}{M!}g_{\alpha \beta_1,...,\beta_M}g_{\alpha \gamma_1,...,\gamma_M}\vec{x}_{\beta_1}\cdot \vec{x}_{\gamma_1}...\vec{x}_{\beta_M}\cdot \vec{x}_{\gamma_M}},
\end{equation}
\[
g_{\alpha \alpha_1,...,\alpha_M}:=\int{Y_{\alpha}\epsilon^{a_1...a_M}\partial_{a_1}Y_{\alpha_1}...\partial_{a_M}Y_{\alpha_M}d^{M}\varphi}.
\]

Due to $\zeta$ satisfying (and, apart from $\zeta_0$, being reconstructed from) 
\begin{equation}
\partial_a \zeta=\frac{\vec{p}\partial_a \vec{x}}{\eta \rho}
\end{equation}
($\rho$ being a non-dynamical density on $\Sigma_0^{M}$ of unit weight: $\int{\rho d^{M}\varphi}=1$)
\begin{equation}
L_{\alpha}:=\frac{1}{\mu_{\alpha}}\int{\nabla^a Y_{\alpha}\frac{\vec{p}}{\rho}\partial_a \vec{x}}
\label{vnpdgen}
\end{equation}
-which is of the form $\int{f^a}\vec{p}\partial_a\vec{x}d^M\varphi$ with $\nabla_a f^a=-Y_{\alpha}$ (hence a complete set of generators of volume non-preserving diffeomorphisms) is (up to the constant factor $\eta$) identical to the $\alpha$-component of $\zeta$.
As shown in \cite{model}, the Poisson-bracket (if quantized: commutator) of the generators (\ref{vnpdgen}) is ($\left[\alpha,\alpha^{\prime} \right] $ denoting antisymmetrization)
\begin{equation}
[L_{\alpha},L_{{\alpha}^{\prime}}]\approx e_{[\alpha,{\alpha}^{\prime}]\epsilon}L_{\epsilon}
\end{equation}
with 
\begin{equation}
e_{\alpha \beta \gamma}:=\frac{\mu_{\beta}-\mu_{\gamma}}{\mu_{\alpha}}d_{\alpha \beta \gamma}, \ d_{\alpha \beta \gamma}:=\int{Y_{\alpha} Y_{\beta} Y_{\gamma}\rho d^M \varphi},
\end{equation}
$\approx$ meaning 'modulo VPD'.\\
In this note we will calculate explicitely the structure constants of the Lie algebra of vector fields on a 2-dimensional compact orientable manifold, using as a basis the generators $L_{\alpha}$, as well as $\phi_{\alpha}:=g_{\alpha \mu \nu} \vec{x}_{\mu}\vec{p}_{\nu}$ and (absent on $S^2$) generators $H_{r}:=\int{h_{(r)}^a\vec{p}\partial_a\vec{x}}d^2\varphi$ corresponding to harmonic vector fields ($\nabla_a h^a_{(r)} = 0$, $\epsilon_{ab} \nabla^a h^b_{(r)}=0$, $ r=1,...,2g$, where g is the genus of the manifold), satisfying
\begin{equation}
\begin{matrix}
\left[\phi_{\alpha},\phi_{{\alpha}^{\prime}}\right]=g_{\alpha {\alpha}^{\prime} \epsilon} \phi_{\epsilon}& \\
\left[H_{r},H_{r^{\prime}}\right]=g_{r r^{\prime}\epsilon} \phi_{\epsilon} &\\
\left[\phi_{\alpha}, H_{r}\right]=g_{\alpha r  \epsilon}\phi_{\epsilon}&\\
\left[L_{\alpha},L_{{\alpha}^{\prime}}\right]=e_{[\alpha,{\alpha}^{\prime}]\epsilon}L_{\epsilon}+\tilde{g}_{\alpha{\alpha}^{\prime}\epsilon}\phi_{\epsilon}+k_{\alpha \alpha^{\prime}r}H_{r}&\\
\left[L_{\alpha},\phi_{{\alpha}^{\prime}}\right]=g_{\alpha \alpha^{\prime} \epsilon} L_{\epsilon}+\tilde{e}_{\alpha {\alpha}^{\prime} \epsilon} \phi_{\epsilon}+c_{\alpha \alpha^{\prime} r}H_{r}&\\
\left[L_{\alpha},H_{r}\right]=g_{\alpha r \epsilon }L_{\epsilon}+\tilde{c}_{\alpha r \epsilon}\phi_{\epsilon}+y_{\alpha r r^{\prime} }H_{r^{\prime}}&
\end{matrix}
\end{equation}
While the non-trivial fact that the $L_{\epsilon}$ component of $\left[L_{\alpha},\phi_{\alpha^{\prime}}\right]$ is $g_{\alpha \alpha^{\prime} \epsilon}$, i.e. that (modulo VPD) $\zeta$ transforms as a function under VPD (implying in particular that $\int{\vec{p}\zeta d^{M}\varphi} $ weakly commutes with the $\phi_{\alpha}$) was already noted in \cite{modinv}, we will (with $\hat{\epsilon}_{ab}:=\rho \epsilon_{ab}$ and $\int f := \int f \rho d^2\varphi$) now prove that

\normalfont
\begin{subequations}
\begin{align}
g_{r r^{\prime}\epsilon}=\int{\hat{\epsilon}_{ab}h_{(r)}^ah_{(r^{\prime})}^bY_{\epsilon}}
\label{th1}\\
g_{ \alpha r   \epsilon}=\int{Y_{\alpha} h_{(r)}^a\nabla_{a}Y_{\epsilon}}
\label{th2}\\
\tilde{g}_{\alpha{\alpha}^{\prime}\epsilon}=\frac{\mu_{\epsilon}-\mu_{\alpha}-\mu_{\alpha}^{\prime}}{\mu_{\alpha}\mu_{{\alpha}^{\prime}}\mu_{\epsilon}}g_{\alpha {\alpha}^{\prime} \epsilon}
\label{th3}\\
k_{ \alpha  \epsilon r}=(\frac{1}{\mu_{\alpha}}+\frac{1}{\mu_{\epsilon}})g_{\alpha r   \epsilon}
\label{th4}\\
\tilde{e}_{\alpha {\alpha}^{\prime} \epsilon}=d_{\alpha {\alpha}^{\prime} \epsilon}+\frac{1}{2}(e_{\alpha {\alpha}^{\prime} \epsilon}-e_{\epsilon \alpha {\alpha}^{\prime} })
\label{th5}\\
c_{\alpha \alpha^{\prime} r}=\int Y_{\alpha}   \hat{\epsilon}^{ab} \nabla_a  Y_{\alpha^{\prime}} h_{(r)b}
\label{th6}\\
\tilde{c}_{\alpha r \epsilon}=(\frac{1}{\mu_{\alpha}}-\frac{1}{\mu_{\epsilon}})c_{ \epsilon \alpha  r}
\label{th7}\\
y_{\alpha r r^{\prime} }=\int Y_{\alpha} h_{(r)}^c h_{(r^{\prime})c}
\label{th8}
\end{align}
\end{subequations}

\textbf{(\ref{th1})}
Integrating by parts with respect to $\nabla_a$ and using $\nabla_a h_{(r)}^a=0$ we have
\begin{eqnarray}
\left[H_{r},H_{r^{\prime}}\right]=\int (h_{(r)}^a\nabla_a h_{(r^{\prime})}^b-(r \leftrightarrow r^{\prime}))\vec{p}\partial_b \vec{x}d^2\varphi &\nonumber\\
=-\int \underbrace{(h_{(r)}^a h_{(r^{\prime})}^b-(r \leftrightarrow r^{\prime}))}_{\epsilon^{ab}\epsilon_{cd} h_{(r)}^c  h_{(r^{\prime})}^d}\nabla_a(\vec{\frac{p}{\rho}}\partial_b \vec{x})=-\int \hat{\epsilon}_{cd}h_{(r)}^c h_{(r^{\prime})}^d\underbrace{\left\{\frac{\vec{p}}{\rho},\vec{x}\right\}}_{\vec{p}_{\alpha}\cdot\vec{x}_{\beta}g_{\alpha \beta \epsilon} Y_{\epsilon}}\nonumber\\
=\phi_{\epsilon} \int \hat{\epsilon}_{cd}h_{(r)}^c h_{(r^{\prime})}^d Y_{\epsilon}
\end{eqnarray}

\textbf{(\ref{th2})}
After decomposing $\vec{p}$ and $\vec{x}$ into Laplace-eigenfunctions, and using the fact that divergence free
vectorfields form a subalgebra, the commutator $\left[ H_{r},\phi_{\alpha}\right]$ takes the following form
\begin{equation}
\vec{p}_{\beta} \cdot \vec{x}_{\gamma}\int(\hat{\epsilon}^{ac}\nabla_b \partial_a Y_{\alpha} h_{(r)}^b-\hat{\epsilon}^{ab}\partial_a Y_{\alpha} \nabla_b h_{(r)}^c)Y_{\beta} \partial_c Y_{\gamma}
\end{equation}
Integrating the first term by parts w.r.t. $\nabla_a$ and using $\nabla_a h_{(r)}^a=0$ gives (\ref{th2})
\begin{eqnarray}
-\vec{p}_{\beta} \cdot \vec{x}_{\gamma}\int\nabla_b Y_{\alpha} h_{(r)}^b\underbrace{\hat{\epsilon}^{ac}\partial_a Y_{\beta} \partial_c Y_{\gamma}}_{g_{\beta \gamma \epsilon} Y_{\epsilon}}\nonumber\\
- \vec{p}_{\beta} \cdot \vec{x}_{\gamma}\int(\hat{\epsilon}^{ab}\nabla_b h_{(r)}^c) \underbrace{(\partial_a Y_{\alpha}\partial_c Y_{\gamma} -(a\leftrightarrow c))}_{\epsilon_{ac}\left\{Y_{\alpha},Y_{\gamma}\right\}\rho}Y_{\beta}\nonumber\\
=\phi_{\epsilon}\int{h_{(r)}^a\nabla_{a}Y_{\alpha}Y_{\epsilon}}
\end{eqnarray}

\textbf{(\ref{th3})}
In order to find the coefficients of $\phi_{\epsilon}$ in 
\begin{equation}
\left[L_{\alpha},L_{{\alpha}^{\prime}}\right]=\frac{1}{\mu_{\alpha} \mu_{\alpha^{\prime}}}\int{(\nabla^b Y_{\alpha}\nabla_{b}(\nabla^{a}Y_{\alpha^{\prime}})-(\alpha\leftrightarrow\alpha^{\prime}))\frac{\vec{p}}{\rho}\partial_a\vec{x}}
\end{equation}
we insert the projector onto divergence-free vector fields (used in \cite{gold2}, in the mid-eighties, in the proof of classical Lorentz invariance of
the light-cone gauge-fixed description of
bosonic M-branes)
\begin{eqnarray}
F^a_c(\varphi,\tilde{\varphi}):=
-\sum_{\epsilon}\frac{1}{\mu_{\epsilon}}\partial_c Y_{\epsilon}(\varphi)\tilde{\partial}^a Y_{\epsilon}(\tilde{\varphi})+\frac{1}{\rho}\delta_c^a \delta(\varphi,\tilde{\varphi})\nonumber\\
=\sum_{\epsilon}\frac{1}{\mu_{\epsilon}} \hat{\epsilon}_{c c^{\prime}}\partial^{c^{\prime}}Y_{\epsilon}(\varphi) \hat{\epsilon}^{a a^{\prime}}\tilde{\partial}_{a^{\prime}} Y_{\epsilon}(\tilde{\varphi})+\sum_{r}h_{(r)c}(\varphi)h_{(r)}^a(\tilde{\varphi})
\label{projektor}
\end{eqnarray}
As 
\begin{eqnarray}
\int{\hat{\epsilon}^{a a^{\prime}}\tilde{\nabla}_{a^{\prime}}Y_{\epsilon}(\widetilde{\varphi})\widetilde{\vec{p}\partial_a \vec{x}}}d^2\tilde{\varphi}
=-\int Y_{\epsilon} \underbrace{\hat{\epsilon}^{a a^{\prime}}\partial_{a}\vec{x}\partial_{a^{\prime}}\frac{\vec{p}}{\rho}}_{=:\left\{\vec{x},\frac{\vec{p}}{\rho}\right\}}\nonumber\\
=-g_{\epsilon \mu \nu}\vec{x}_{\mu}\vec{p}_{\nu}=-\phi_{\epsilon}
\end{eqnarray}
one has  (in the last step integrating by parts with respect to the inner $\nabla_b$)
\begin{eqnarray}
\tilde{g}_{\alpha \alpha^{\prime}\epsilon}=- \frac{1}{\mu_{\alpha} \mu_{\alpha^{\prime}}\mu_{\epsilon}}\int{(\nabla^b Y_{\alpha}\nabla_{b}(\nabla^{c}Y_{\alpha^{\prime}})-(\alpha\leftrightarrow\alpha^{\prime}))\hat{\epsilon}_{c c^\prime}\nabla^{c^{\prime}}Y_{\epsilon}}\nonumber\\
=-(\frac{1}{\mu_{\alpha^{\prime}}}+\frac{1}{\mu_{\alpha}})\frac{1}{\mu_{\epsilon}}\int{\underbrace{Y_{\alpha}\hat{\epsilon}_{c c^{\prime}}\nabla^c Y_{\alpha^{\prime}}\nabla^{c^{\prime}}Y_{\epsilon}}_{=g_{\alpha \alpha^{\prime}\epsilon}}}\nonumber\\
+\frac{1}{\mu_{\alpha}\mu_{\alpha^{\prime}}\mu_{\epsilon}}\int{\underbrace{(\nabla^{b}Y_{\alpha}\nabla^c Y_{\alpha^{\prime}}-(\alpha \leftrightarrow \alpha^{\prime}))}_{\left\{Y_{\alpha},Y_{\alpha^{\prime}}\right\}\hat{\epsilon}^{bc}}\hat{\epsilon}_{c c^{\prime}}\nabla_b \nabla^{c^{\prime}}}Y_{\epsilon}
\end{eqnarray}
which immediately yields (\ref{th3}).

\textbf{(\ref{th4})}
Similarly, the coefficient of $H_r$ is (taking into account the harmonic part of the projector (\ref{projektor}) and integrating by parts with respect to the inner $\nabla^b$ ) 
\begin{eqnarray}
\frac{1}{\mu_{\alpha} \mu_{\alpha^{\prime}}}\int(\mu_{\alpha} \nabla^{a}Y_{\alpha^{\prime}}Y_{\alpha}-(\alpha\leftrightarrow\alpha^{\prime}))h_{(r)a}(\varphi)\nonumber\\
-\frac{1}{\mu_{\alpha} \mu_{\alpha^{\prime}}}\int \underbrace{(\nabla^b Y_{\alpha} \nabla^{a}Y_{\alpha^{\prime}}-(\alpha\leftrightarrow\alpha^{\prime}))}_{\hat{\epsilon}^{ba} \left\{Y_{\alpha},Y_{\alpha^{\prime}}\right\}}\nabla_b h_{(r)a}(\varphi)\nonumber\\
= (\frac{1}{\mu_{\alpha}}+\frac{1}{\mu_{\alpha^{\prime}}})g_{\alpha  r \alpha^{\prime}}
\end{eqnarray}
The second term vanishes due to $\epsilon^{ba} \nabla_b h_{(r)a}=0$, while the first one gives the desired result.\\

\textbf{(\ref{th5})}
In order to decompose 
\begin{eqnarray}
\left[L_{\alpha}, \phi_{\delta}\right]=\int{\frac{1}{\mu_{\alpha}}\left[\hat{\epsilon}^{ac}\nabla_cY_{\delta}\nabla_a \nabla^b Y_{\alpha}-\nabla^a Y_{\alpha} \hat{\epsilon}^{bc} \nabla_a \nabla_ c Y_{\delta}\right]\vec{p}\partial_b \vec{x}}d^2 \varphi\nonumber\\
\equiv \int{f^d \vec{p}\partial_d \vec{x}d^2\varphi}
\end{eqnarray}
one simply inserts the completeness relation for vector fields, cp. (\ref{projektor}), obtaining 
\begin{eqnarray}
\int\int\frac{f^c(\varphi)}{\mu_{\epsilon}}\left[\partial_c Y_{\epsilon}(\varphi)\tilde{\partial}^{d}Y_{\epsilon}(\tilde{\varphi})+\mu_{\epsilon} \sum_{r}h_{(r)c}(\varphi)h_{(r)}^d(\tilde{\varphi})\right]\frac{\tilde{\vec{p}}}{\tilde{\rho}}\widetilde{\partial_{d}\vec{x}} \nonumber \\
+\int \int \frac{f^c(\varphi)}{\mu_{\epsilon}}\left[\epsilon_{c c^{\prime}}\partial^{c^{\prime}}Y_{\epsilon}(\varphi)\epsilon^{dd^{\prime}}\partial_{d^{\prime}}Y_{\epsilon}(\tilde{\varphi})\right]\frac{\tilde{\vec{p}}}{\tilde{\rho}}\widetilde{\partial_{d}\vec{x}}
\end{eqnarray}
So the coefficient of $\phi_{\epsilon}$ is 
\begin{eqnarray}
\frac{1}{\mu_{\alpha} \mu_{\epsilon}}\int (\partial^a Y_{\epsilon}\underbrace{\left[\nabla^b Y_{\alpha}\nabla_b \nabla_a Y_{\delta}+\nabla^b Y_{\delta}\nabla_a \nabla_b Y_{\alpha}\right]}_{\nabla_a\left[...\right]} -\nabla^b Y_{\epsilon}\nabla_b Y_{\delta}\Delta Y_{\alpha})\nonumber\\
=\frac{1}{\mu_{\alpha}}\int \nabla^b Y_{\alpha}Y_{\epsilon}\nabla_b Y_{\delta}+ \frac{1}{\mu_{\epsilon}}\int\nabla^b Y_{\epsilon}Y_{\alpha}\nabla_b Y_{\delta}\nonumber\\
=d_{\alpha \delta \epsilon} +\frac{1}{2}(e_{\alpha \delta \epsilon}-e_{\epsilon \alpha \delta})
\end{eqnarray}

And the coefficient of $L_{\epsilon}$ 

\begin{eqnarray}
 \frac{1}{\mu_{\alpha}}\int Y_{\epsilon} ( \nabla_c \nabla^b Y_{\alpha} \hat{\epsilon^{ca}} \nabla_b \nabla_a Y_{\delta}+\underbrace{\nabla^b Y_{\alpha}\hat{\epsilon}^{c a}\nabla_c \nabla_b \nabla_a Y_{\delta}}_{\hat{\epsilon}^{ac}\nabla^b Y_{\alpha}R_{adcb} \nabla^d Y_{\delta}} \nonumber\\
-\hat{\epsilon}^{ac}\nabla_a \nabla^b Y_{\alpha}\nabla_b\nabla_c Y_{\delta} -\underbrace{\nabla_b \nabla_a \nabla^b Y_{\alpha} \hat{\epsilon}^{ac}\nabla_c Y_{\delta} }_{(-R_{ab} \nabla^b + \nabla_a \Delta) Y_{\alpha} \hat{\epsilon}^{ac} \nabla_c Y_{\delta}})=g_{\alpha \delta \epsilon}
\label{l}
\end{eqnarray}
where we used the simple form of the Riemann tensor in two dimensions $R_{abcd}=K (g_{ac}g_{bd}-g_{ad}g_{bc})$, where K is the Gaussian curvature; so the second term is canceled by the $R_{ab}$ part of the last term, the first term is cancelled by the third one and the remaining part is exactly $g_{\alpha \delta \epsilon}$.\\

\textbf{(\ref{th6})}
The coefficient of $H_r$ in $\left[L_{\alpha},\phi_{\delta}\right]$ is (integrating both terms by parts w.r.t. $\nabla_a$ and using that $\epsilon^{ab} \nabla_a h_{(r)b}=0$)
\begin{eqnarray}
\int f^b h_{(r)b}=\frac{1}{\mu_{\alpha}}\int (\hat{\epsilon}^{ac}\nabla_c Y_{\delta} \nabla_a \nabla^b Y_{\alpha}-\hat{\epsilon}^{bc}\nabla^a Y_{\alpha}\nabla_a \nabla_c Y_{\delta})h_{(r)b} \nonumber\\
= -\int \hat{\epsilon}^{ba} h_{(r)a} \nabla_b Y_{\alpha} Y_{\delta}
\end{eqnarray}

\textbf{(\ref{th7},\ref{th8})}
In order to determine the coefficients of $L_{\epsilon}, \phi_{\epsilon}$ and $H_{r^{\prime}}$ in $\left[L_{\alpha}, H_r\right]$ we do the following
\begin{eqnarray}
\left[L_{\alpha}, H_r\right]=\frac{\vec{p}_{\beta} \cdot \vec{x}_{\gamma}}{\mu_{\alpha}} \int (\nabla^a Y_{\alpha} \nabla_a h^{(r)b}-h^{(r)a}\nabla_a \nabla^b Y_{\alpha})Y_{\beta}\partial_b Y_{\gamma}\nonumber\\
=\underbrace{-\frac{\vec{p}_{\beta} \cdot \vec{x}_{\gamma}}{\mu_{\alpha}} \int\underbrace{(\nabla^a Y_{\alpha}h^{(r)b}-h^{(r)a}\nabla^bY_{\alpha})}_{\hat{\epsilon}^{ab}\hat{\epsilon}^{cd}\nabla_c Y_{\alpha}h^{(r)}_d}\nabla_a(Y_{\beta} \partial_b Y_{\gamma})}_{\frac{1}{\mu_{\alpha}}c_{ \epsilon \alpha r} \phi_{\epsilon}}\nonumber\\
-\frac{1}{\mu_{\alpha}}\int \Delta Y_{\alpha} h_{(r)}^b\frac{\vec{p}}{\rho}\partial_b \vec{x}
\end{eqnarray}
where we integrated by parts w.r.t. $\nabla_a$.
The first term gives a contribution to the coefficient of $\phi_{\epsilon}$ directly, while the second one necessitates a bit more work. After inserting the completeness relation (\ref{projektor}) we are left with three terms proportional to $L_{\epsilon}, \phi_{\epsilon}$ and $H_r$ 
\begin{eqnarray}
\underbrace{\int Y_{\alpha} h_{(r)}^c\partial_cY_{\epsilon}}_{g_{\alpha r \epsilon }}L_{\epsilon}\\
\underbrace{-\frac{1}{\mu_{\epsilon}}\int Y_{\alpha} h_{(r)}^c\epsilon_{c c^{\prime}}\partial^{c^{\prime}}Y_{\epsilon}}_{-\frac{1}{\mu_{\epsilon}}c_{ \epsilon \alpha r}}\phi_{\epsilon}\\
\underbrace{\int Y_{\alpha} h_{(r)}^c h_{(r^{\prime})c}}_{y_{\alpha r r^{\prime}}}H_{r^{\prime}}
\end{eqnarray}

\bfseries Acknowledgments \normalfont We thank the Swedish Research Council for financial support, as well as M.Bordemann, J.Mickelsson and P.Michor for discussions and correspondence.

\end{document}